\documentclass[a4paper,10pt,twoside]{cpc-hepnp}

\usepackage{multicol}
\usepackage{graphicx}
\usepackage{booktabs}
\usepackage{amssymb,bm,mathrsfs,bbm,amscd}
\usepackage[tbtags]{amsmath}
\usepackage{lastpage}

\begin{document}

\fancyhead[c]{\small Chinese Physics C~~~Vol. XX, No. X (201X)
XXXXXX} \fancyfoot[C]{\small 010201-\thepage}

\footnotetext[0]{Received XX September 2014}

\title{Induced radioactivity analysis for the NSRL Linac in China using Monte Carlo simulations and gamma-spectroscopy\thanks{Supported by National Natural Science Foundation of China (11175180 and 11175182) }}

\author{%
      He Lijuan(何丽娟)$^{1;1)}$\email{juan@mail.ustc.edu.cn}%
\quad Li Yuxiong(李裕熊)$^{1;2)}$\email{lyx@ustc.edu.cn}%
\quad Li Weimin(李为民)$^{1}$
\quad Chen Zhi(陈志)$^{2}$\\
\quad Chen Yukai(陈裕凯)$^{1}$
\quad Ren Guangyi(任广益)$^{1}$
}
\maketitle

\address{%
$^1$ National Synchrotron Radiation Laboratory, University of Science and Technology of China, Hefei, Anhui 230029, P.R.China\\
$^2$ School of Nuclear Science and Technology of USTC, No.443, Huangshan Road, Hefei, Anhui 230029, P.R.China\\
}

\begin{abstract}
The 200-MeV electron linac of the National Synchrotron Radiation Laboratory (NSRL) located in Hefei is one of the earliest high-energy electron linear accelerators in China. The electrons are accelerated to 200 MeV by five acceleration tubes and are collimated by scrapers. The scraper aperture is smaller than the acceleration tube one, so some electrons hit the materials when passing through them. These lost electrons cause induced radioactivity mainly due to bremsstrahlung and photonuclear reaction. This paper describes a study of induced radioactivity for the NSRL Linac using FLUKA simulations and gamma-spectroscopy. The measurements showed that electrons were lost mainly at the scraper. So the induced radioactivity of the NSRL Linac is mainly produced here. The radionuclide types were simulated using the FLUKA Monte Carlo code and the results were compared against measurements made with a High Purity Germanium (HPGe) gamma spectrometer. The NSRL linac had been retired because of upgrading last year. The removed components were used to study induced radioactivity. The radionuclides confirmed by the measurement are: $^{57}$Ni, $^{52}$Mn, $^{51}$Cr, $^{58}$Co, $^{56}$Co, $^{57}$Co, $^{54}$Mn, $^{60}$Co and $^{22}$Na, the first eight nuclides of which are predicted by FLUKA simulation. The research will provide the theoretical basis for the similar accelerator decommissioning plan, and is significant for accelerator structure design, material selection and radiation protection design.
\end{abstract}



\footnotetext[0]{\hspace*{-3mm}\raisebox{0.3ex}{$\scriptstyle\copyright$}2013
Chinese Physical Society and the Institute of High Energy Physics
of the Chinese Academy of Sciences and the Institute
of Modern Physics of the Chinese Academy of Sciences and IOP Publishing Ltd}%

\begin{multicols}{2}

\section{Introduction}

The Synchrotron Radiation Facility of the National Synchrotron Radiation Laboratory (NSRL) located in Hefei, China consists of a 200-MeV electron linear accelerator and an 800-MeV electronic storage ring\cite{lab1}. Constructed in 1989, the 200-MeV Electron linac (Fig.~\ref{fig1}) is one of the earliest high-energy electron linear accelerators in China. Lower-energy electrons emitted from the accelerator electron gun are progressively accelerated to about 200 MeV in five acceleration tubes, and then are injected into the electron storage ring, reaching the final energy of 800 MeV. The linear accelerator is a traveling wave linear accelerator, consisting of pre-injection (including electron gun, pre-buncher, buncher and a 3-m acceleration segment), four 6-m acceleration segment, some beam sensing elements, and focusing inserts. The accelerator's total length is 35 m. The electrons are accelerated in the acceleration tube to 32 MeV, 74 MeV, 116 MeV, 158 MeV, respectively, and the electronic energy reaches 200 MeV at the end of the acceleration trip. A scraper is used for beam collimation and is located at the end of each acceleration tube, with a length that increases with the energy to provide sufficient collimation.
\begin{center}
\includegraphics[width=7cm,height=4.5cm]{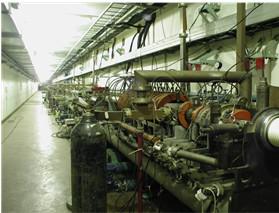}
\figcaption{\label{fig1}   The NSRL Linac in 2012 before it was retired. }
\end{center}

The NSRL Linac was retired in 2012 due to a plan to upgrade the facility. The decommissioning produced many real challenges related to the methods to handle the large size of the device and the time required to store structural materials until the radioactivity is reduced in the waste repository? These problems should be solved as soon as possible in order for the upgrade to proceed smoothly. At present, no prior decommissioning experience existed in China on such a high-energy electron linear accelerator. Therefore, the decommission project provided an opportunity to study and evaluate the challenges, in particular, involving the induced radioactivity for the purposes of radiation protection.

For a well-shielded accelerator facility, the induced radioactivity is an important source of radiation for people who perform the decommissioning. In addition, information on the level of induced radioactivity is needed for disposing of the retired accelerator materials. Therefore it is significant to predict such radioactivity level accurately.

Induced radioactivity for other types of accelerators has been investigated and was found to contain high levels. Evdokimoff et al.\cite{lab2} studied induced radioactivity for a medical accelerator operated in 18-MeV photon mode for radionuclidic exposure. Most activated components appeared to decay with a short-life time (about several minutes). Calandrino et al.\cite{lab3} described the decommissioning of a compact, self-shielded, medical cyclotron (whose energy is 11 MeV). In their study, a Monte Carlo simulation of the possible nuclear reactions (proton and neutron activation) was performed. Meanwhile, residual activities were analyzed with a Ge(Li) detector and compared with simulation data. Its decommissioning did not represent a risk for the involved staff, but due to the presence of long-lived radioisotopes (such as $^{56}$Fe, $^{60}$Co, $^{63}$Ni, etc), the components were to be treated as low level radioactive waste and stored.

Mora et al.\cite{lab4} studied the decommissioning of a 400-MeV Linac, which was commissioned in 1969 and dismantled in 1993. The scraps had been temporarily stored, waiting for definitive disposal. The simulation results showed that $^{22}$Na, $^{54}$Mn, $^{57}$Co and $^{60}$Co were expected activation isotopes in the materials. And the experimental results were in good agreement with other studies.

Rokni et al.\cite{lab5} described the induced radioactivity of several materials, which were irradiated in the stray radiation field generated by the interaction of a 28.5 GeV electron beam. The isotopes generated in the copper sample have: $^{46}$Sc, $^{54}$Mn, $^{59}$Fe, $^{56}$Co, $^{57}$Co, $^{58}$Co, $^{60}$Co and $^{65}$Zn. The specific activity induced in it was measured by gamma spectroscopy and calculated using FLUKA. The calculated activities are compared to the experimental values and differences are discussed.

Sheu et al.\cite{lab6} investigated induced radioactivity due to the operations of a 3-GeV electron accelerator at the Taiwan Photon Source. Fasso et al.\cite{lab7} predicted induced radioactivity for SLAC. Ni et al.\cite{lab8} analyzed photon activation for river sample using a 60-MeV linear electron accelerator.

The radiation protection group of NSRL has published many papers about induced radioactivity of NSRL. Li et al.\cite{lab9} analyzed and discussed the data and spectrum acquired from the induced radiation field and spectrum measurement. Xu et al.\cite{lab10} analyzed the dust activation status at different positions in NSRL 200 MeV linac tunnel. Xu et al.\cite{lab11} introduced the kind of main radioactive nuclides produced by dust via photonuclear reaction experiment. Hong et al.\cite{lab12} studied the potential radiological effect of induced radionuclides in the soil surrounding NSRL on ground water.

However, the induced radioactivity study for such type high-energy electron linear accelerator is very seldom. Particle accelerators that operate at energy greater than 10 MeV will produce induced radioactivity due to well-known photonuclear interactions\cite{lab13}. However, the assessment of such as induced radioactivity of large-scale installations is a time-consuming process. Experimental methods are time-consuming and require special radiation safety measures to protection the investigators. On the other hand, empirical formula cannot yield sufficiently accurate results. Therefore, Monte Carlo simulation method is the primary means to study such problems. In the decommissioning process of the NSRL Linac, the induced radioactivity assessment will be based on Monte Carlo simulation method as the main auxiliary means to expand a series of studies. Because the scraper aperture is smaller than the acceleration tube one, it is reasonable to assume that particles mainly were lost at the scraper. At the same time, the measurements show that the induced radioactivity is due mainly to the scraper and the rest can be ignored. This paper describes the Monte Carlo study of the induced radioactivity in the scraper.

\section{Materials and methods}

\subsection{Theory}

Two types of radiation hazards produced by accelerator need to be considered: prompt radiation and delayed radiation which is also called induced radioactivity. The prompt radiation is generated only when the accelerator is running, and is easily recognized and shielded. The induced radioactivity, on the other hand, exists even when the power is off. Compared with the prompt radiation, the induced radioactivity is more covert, because the hazard can pose a real threat. Therefore the induced radioactivity is the primary source of external irradiation for the staff who works in the well-shielded high-energy accelerator environment\cite{lab9}. For the retired NSRL Linac, the induced radioactivity research is the focus of the work.

When the power is on, electrons emitted from the electron gun are accelerated and transported in the vacuum tube. Due to energy dispersion, some electrons will inevitably impact on the structural materials, causing the induced radiation field. Unlike the proton and ion accelerators, the induced radioactivity generated from an electron accelerator is not dominated by interaction between the original particles (electrons) and the medium. The reason is that the cross section of nuclear reaction is minimal for electron with any energy\cite{lab14}. Its mechanism is: ⑴ the energy of electrons disappear due to bremsstrahlung, ⑵ high-energy photons (higher than 10 MeV) interact with medium, ⑶ subsequent neutrons and pions induce nuclear reactions. Therefore, the radiation field in the high-energy electron linear accelerator generated during operation is a mixed radiation field, composed of electrons, photons and neutrons. These radiations may irradiate the human body. In addition, the high-energy photons and neutrons may also interact with the materials in the radiation field. The interactions may activate these substances via photon and neutron activation processes, leading to a variety of radionuclides\cite{lab15,lab16}. In general, the new nuclides can be classified into two categories\cite{lab17}: ⑴ relatively shorter lived, neutron-deficient nuclides generated through photonuclear reaction, such as (γ,n), (γ,2n), (γ,3n), (γ,p) and (γ,np). ⑵ relatively longer lived and thus more hazardous radionuclides generated due to neutron activation. However, the neutron field produced in the NSRL Linac is an indirect neutron field. The substances activated by neutron are therefore fewer per unit time being exposed to it. Meanwhile, the neutrons generated in the electron accelerator (within 200 MeV), the energy of which is lower, are not enough to produce stronger neutron activation. At the present, radionuclides measured in the experiment are generated by the following photonuclear reaction: (γ,n), (γ,p), (γ,2n) and so on, and neutron activation can be ignored.

\subsection{Investigation of the induced radioactivity}

The 200-MeV electron linear accelerator of NSRL consists of five acceleration tubes, and each tube is followed by a scraper, the length of which increases slightly with the energy. The electrons are accelerated up to the expected energy in each acceleration tube and collimated by the scraper followed each acceleration tube. The radius of the scraper is 4.8 cm, and the bore diameter is 1 cm. The pore size of the disk-loaded waveguide is 2.1977 cm. The scraper aperture is smaller than the disk-loaded waveguide one, so some electrons will hit the scraper when passing through them due to energy dispersion. These lost electrons will cause induced radioactivity via photonuclear reaction. The measurements showed that electrons were lost mainly at the scraper during the accelerating period. So the induced radioactivity at the scraper is the most serious in the NSRL Linac and other structures can be ignored as indicated before already.

The absorber material of the scraper is red copper (type T2) whose elemental composition is shown in Table~\ref{tab1}. Through the investigation of the related literatures\cite{lab7,lab17,lab18,lab19}, radionuclides which may be generated at the scraper were preliminarily determined.
\end{multicols}

\begin{center}
\tabcaption{ \label{tab1}  Elemental composition of the red copper$^{a}$.}
\footnotesize
\begin{tabular*}{178mm}{@{\extracolsep{\fill}}ccccccccc}
\toprule Symbols of element & Cu  & Ag & O & S &  Fe   & Ni & Zn & Pb\\
\hline
Element content (\%)\hphantom{00} & \hphantom{0}99.90 & \hphantom{0}--- & $\leq$0.06 & $\leq$0.005 & $\leq$0.005 & $\leq$0.005 & $\leq$0.005 & $\leq$0.005\\
Relative atomic mass\hphantom{00} & \hphantom{0}63/65 & \hphantom{0}107/109 & 16/17/18 & 32/33 & 54/56 & 58/60/61 & 64/66/67 & 204/206\\
\hphantom{00} & \hphantom{0} & \hphantom{0} &  & 34/36 & 57/58 & 62/64 & 68/70 & 207/208\\
\bottomrule
\end{tabular*}%
\end{center}
$^{a}$The natural abundance of the element in Table~\ref{tab1} is not equal to zero. The element whose content is less than o.oo5\% hasn't been given.

\begin{multicols}{2}

\subsection{Monte Carlo simulations using the FLUKA code}

In this paper, the fourth scraper was chosen as an example to illustrate the induced radioactivity of the NSRL Linac. The electrons are accelerated to 158 MeV in the fourth acceleration tube, and then collimated in the fourth scraper. The scraper is made of the T2 copper. The actual structure of the device is shown in Fig.~\ref{fig2}. The induced radioactivity of the fourth scraper is simulated by Monte Carlo program FLUKA\cite{lab20,lab21}. FLUKA is now capable of giving estimates of the number and types of radionuclides created by photonuclear reactions at high-energy electron linear accelerator. The initial conditions of simulation are as follows. The electron energy is 158 MeV. The radius of annular beam is 0.5 cm, the direction of which is shown in Fig.~\ref{fig3}. The target is a cylinder made of native copper, with a radius of 4.8 cm and a length of 9 cm. The simplified model is shown in Fig.~\ref{fig3}.The Monte Carlo simulation is one of the important methods for forecasting and analyzing the induced radioactivity. So the radionuclides which may be generated at the scraper were simulated by FLUKA.
\begin{center}
\includegraphics[width=7cm,height=4.5cm]{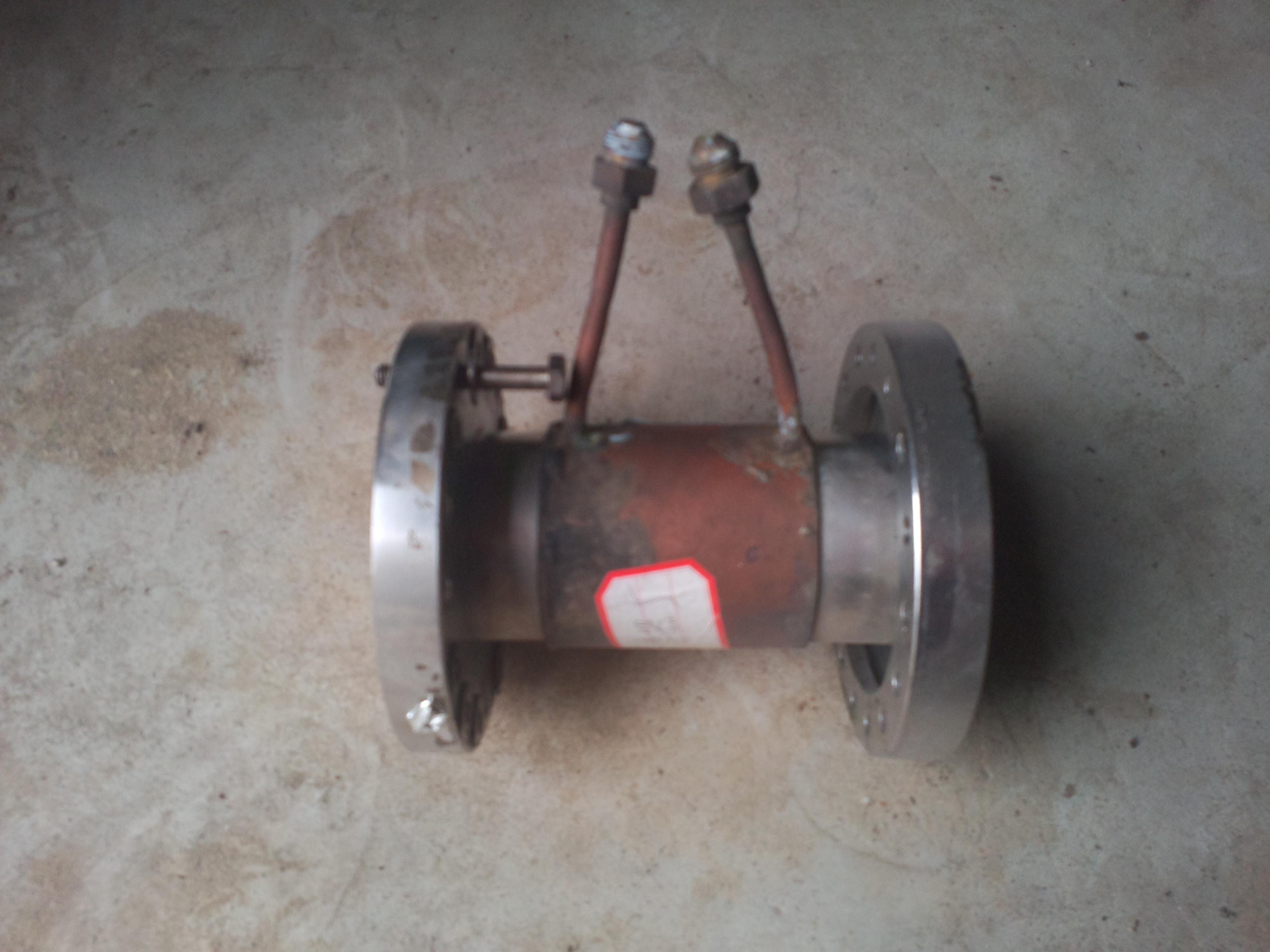}
\figcaption{\label{fig2}   The fourth scraper. }
\end{center}
\begin{center}
\includegraphics[width=7cm,height=4.5cm]{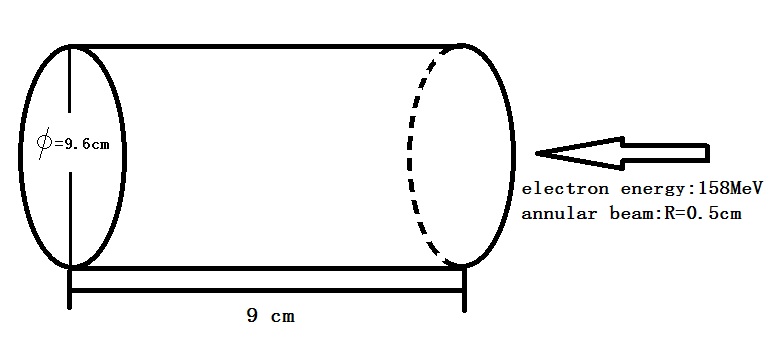}
\figcaption{\label{fig3}   Simplified model of the fourth scraper used for Monte Carlo simulation. }
\end{center}

\subsection{Measurements}

In order to study the induced radioactivity of the scraper, a portable HPGe gamma spectrometer (GR3519 with a relative efficiency of 35\% and an energy resolution of 1.9 keV) made by CANBERRA was used to measure gamma spectra of the scraper followed the fourth acceleration tube (158 MeV) when the power was turned off for one hour and half a month, respectively. The calibration of the HPGe system was performed based on passive efficiency scale division.

\section{Results}

As many as 3,000,000 histories were run in order to obtain reliable results, using 591 minutes on a PC with a Pentium (R) Dual-Core CPU E5500. The estimated relative errors of the FLUKA simulation were below 7\%.

The investigation indicated that the radionuclides generated at the scraper are: $^{56}$Mn, $^{61}$Cu, $^{58}$mCo, $^{64}$Cu, $^{57}$57Ni, $^{52}$Mn, $^{51}$Cr, $^{59}$Fe, $^{58}$Co, $^{46}$Sc, $^{54}$Mn, $^{55}$Fe, $^{60}$Co and $^{63}$Ni. The FLUKA simulation pointed that the radionulides are: $^{62}$Cu, $^{61}$Co, $^{65}$Ni, $^{56}$Mn, $^{61}$Cu, $^{64}$Cu, $^{52}$Mn, $^{48}$V, $^{51}$Cr, $^{59}$Fe, $^{58}$Co, $^{56}$Co, $^{65}$Zn, $^{57}$Co, $^{54}$Mn, $^{49}$V, $^{55}$Fe, $^{60}$Co and $^{63}$Ni. The measurement results were shown in Fig.~\ref{fig4} and Fig.~\ref{fig5}.
\begin{center}
\includegraphics[width=7cm,height=4.5cm]{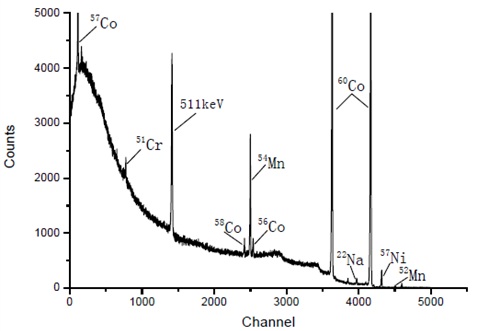}
\figcaption{\label{fig4}    Gamma spectra obtained from the activated fourth scraper for the fourth acceleration tube (158MeV). One hour after the power was off. }
\end{center}
\begin{center}
\includegraphics[width=7cm,height=4.5cm]{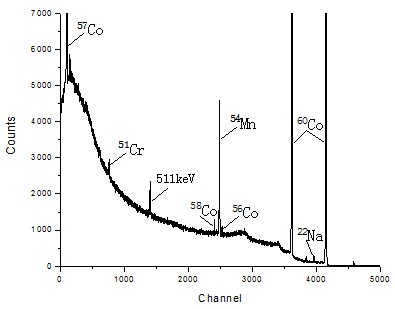}
\figcaption{\label{fig5}   Gamma spectra obtained from the activated fourth scraper for the fourth acceleration tube (158MeV). Half a month after the power was turned off. }
\end{center}

\section{Discussion}

Comparing investigation and simulation results of radionuclides, the nuclides that may be generated in the scraper are listed in Table~\ref{tab2}. The radionuclides which will emit e$^{+}$ are as follows: $^{61}$Cu, $^{64}$Cu, $^{57}$Ni, $^{52}$Mn, $^{58}$Co and $^{56}$Co. Positron-electron annihilation will generate gamma ray carrying the energy of 511 keV. The ray type emitted from 63Ni is beta, which can’t be detected by HPGe gamma spectrometer. The measured radionuclides are also listed in Table~\ref{tab2}, which are compared with the predicted nuclides. This can provide a certain direction for the latter part of the measurements.

By observing Fig.~\ref{fig4}, Fig.~\ref{fig5} and Table~\ref{tab2}, some conclusions can be drawn: ⑴ The induced radionuclides generated in the scraper are mainly $^{57}$Ni, $^{52}$Mn, $^{51}$Cr, $^{58}$Co, $^{56}$Co, $^{57}$Co, $^{54}$Mn, $^{60}$Co, $^{22}$Na, etc, the first eight nuclides of which are predicted. ⑵ Through the comparison of Fig.~\ref{fig4} and Fig.~\ref{fig5}, we can find that the short-life nuclides are markedly reduced, such as $^{57}$Ni and $^{52}$Mn. ⑶ Through the comparison of Fig.~\ref{fig4} and Fig.~\ref{fig5}, we can find that the peak of 511 keV are significantly decreased, indicating that some short-life nuclides which can produce positrons are markedly reduced, such as $^{57}$Ni, $^{52}$Mn. ⑷ The corresponding peaks are not found for some predicted nuclides in the spectrum, and the reasons are as follows: ① Element contents besides $^{63}$Cu and $^{65}$Cu are very low, so the corresponding radionuclides generated by them are few, which are difficult to detect. ② The half-life of some nuclides are very short, so they have decayed to lower limit of measurement when measuring. ③ The energy of the ray emitted by some radionuclides is similar, which is difficult to distinguish in the spectrum. ④ Some radionuclides will emit positrons, then positron-electron annihilation will occur and will produce two gamma photons with 511 keV, but gamma spectrometer cannot distinguish them generated by which nuclide.
\end{multicols}

\begin{center}
\tabcaption{ \label{tab2}  Radionuclides that may be generated in the scraper.}
\footnotesize
\begin{tabular*}{178mm}{@{\extracolsep{\fill}}cccccccc}
\toprule Parent  & Reaction   & Radionuclide & $T_{1/2}$ & Threshold  &  Gamma energy    & Confirmed by  & Confirmed by\\
\ nucleus & type  &  &  & (MeV) &   (keV)   & FLUKA simulation &  HPGe measurement\\
\hline
$^{57}$Fe\hphantom{00} & \hphantom{0}($\gamma$,p) & \hphantom{0}$^{56}$Mn & 2.576h & 10.57 & $\gamma$(846,1810,2113,2522, & Yes & No\\
\hphantom{00} & \hphantom{0} & \hphantom{0} & &  & 2657,2959,3369) &  & \\
$^{63}$Cu\hphantom{00} & \hphantom{0}($\gamma$,2n) & \hphantom{0}$^{61}$Cu & 3.32h & 19.73 & $\gamma$(282,656) & Yes & No\\
$^{63}$Cu\hphantom{00} & \hphantom{0}($\gamma$,sp) & \hphantom{0}$^{58m}$Co & 9.2h & 41.75 & $\gamma$(6.9) & No & No\\
$^{65}$Cu\hphantom{00} & \hphantom{0}($\gamma$,n) & \hphantom{0}$^{64}$Cu & 12.80h & 9.91 & $\gamma$(1345) & Yes & No\\
$^{58}$Ni\hphantom{00} & \hphantom{0}($\gamma$,n) & \hphantom{0}$^{57}$Ni & 36.0h & 12.19 & $\gamma$(127,1377,1757,1919) & No & Yes\\
$^{54}$Fe\hphantom{00} & \hphantom{0}($\gamma$,np) & \hphantom{0}$^{52}$Mn & 5.6d & 20.89 & $\gamma$(744,935,1333,1434) & Yes & Yes\\
$^{54}$Fe\hphantom{00} & \hphantom{0}($\gamma$,sp) & \hphantom{0}$^{51}$Cr & 27.72d & 19.74 & $\gamma$(320) & Yes & Yes\\
$^{58}$Fe\hphantom{00} & \hphantom{0}(n,$\gamma$) & \hphantom{0}$^{59}$Fe & 45.6d & --- & $\gamma$(192,1099,1291) & Yes & No\\
$^{63}$Cu\hphantom{00} & \hphantom{0}($\gamma$,sp) & \hphantom{0}$^{58}$Co & 71.3d & 41.75 & $\gamma$(810,863,1674) & Yes & Yes\\
$^{56}$Ni\hphantom{00} & \hphantom{0}(n,p) & \hphantom{0}$^{56}$Co & 78.76d & --- & $\gamma$(846,1037,1238,1771, & Yes & Yes\\
\hphantom{00} & \hphantom{0} & \hphantom{0} & &  & 2034,2598,3201,3253, &  & \\
\hphantom{00} & \hphantom{0} & \hphantom{0} & &  & 3272,3451,3547) &  & \\
$^{54}$Fe\hphantom{00} & \hphantom{0}($\gamma$,sp) & \hphantom{0}$^{46}$Sc & 83.9d & 37.41 & $\gamma$(889,1120.5) & No & No\\
$^{58}$Ni\hphantom{00} & \hphantom{0}($\gamma$,p) & \hphantom{0}$^{57}$Co & 270.9d & --- & $\gamma$(122,136,692) & Yes & Yes\\
$^{57}$Ni\hphantom{00} & \hphantom{0}(n,p) & \hphantom{0}$^{57}$Co & 270.9d & --- & $\gamma$(122,136,692) & Yes & Yes\\
$^{56}$Fe\hphantom{00} & \hphantom{0}($\gamma$,np) & \hphantom{0}$^{54}$Mn & 303d & 20.42 & $\gamma$(834) & Yes & Yes\\
$^{56}$Fe\hphantom{00} & \hphantom{0}($\gamma$,n) & \hphantom{0}$^{55}$Fe & 2.60a & 11.21 & --- & Yes & No\\
$^{63}$Cu\hphantom{00} & \hphantom{0}($\gamma$,n2p) & \hphantom{0}$^{60}$Co & 5.263a & 18.86 & $\gamma$(1173,1332) & Yes & Yes\\
$^{65}$Cu\hphantom{00} & \hphantom{0}($\gamma$,np) & \hphantom{0}$^{63}$Ni & 92a & 17.11 & --- & Yes & No\\
\bottomrule
\end{tabular*}%
\end{center}

\begin{multicols}{2}

\section{Conclusions}

For the NSRL Linac, the electrons are mainly lost at the scraper, so the induced radioactivity is mostly produced here. In this work, Monte Carlo software FLUKA is used to simulate the induced radioactivity of the scraper. Combined with the research results, the prediction of the radionuclides generated in the scraper is given.

In addition, in order to study the induced radioactivity of the scraper, a portable HPGe gamma spectrometer (GR3519) was used to measure gamma spectra of the scraper followed the fourth acceleration tube (158 MeV) when the power was off for one hour and half a month. By observing the measurement results and the predicted nuclides, we were able to find that only one nuclide was not predicted: $^{22}$Na. This result suggests that the Monte Carlo simulation method was able to forecast the induced radioactivity problem and can also be used in the activation analysis of other high-energy particle accelerators in the future.

Last year, the linac mentioned above was retired due to a planned upgrading. The above work was just the beginning of the decommissioning plan, so there are still many questions to ponder. The future tasks include:

⑴ The radionuclides’ types, content data and decay laws in the scraper need to improve by the long-term measurements. At the same time, in order to determine the focus of each measurement, we need use the radionuclide prediction table as the guidance.

⑵ At present, in order to get the relationship between the radionuclides’ types and the depth, the scraper followed the acceleration tube with 158 MeV has been sliced, in which the radionuclides are being measured. Meanwhile, Monte Carlo program FLUKA is being used in the induced radioactivity simulation.

⑶ In the future, in order to get the corresponding relationship between the electron energy and nuclides’ types, induced radioactivity in the scrapers followed the acceleration tubes with different energy will be simulated and measured.
\\

\acknowledgments{The authors would like to thank the staffs in radiation monitoring stations in Anhui province for their assistance during the measurements. And the authors would like to thank X. George Xu for his assistance in the process of article revision.}

\end{multicols}

\vspace{-1mm}
\centerline{\rule{80mm}{0.1pt}}
\vspace{2mm}

\begin{multicols}{2}

\end{multicols}

\clearpage

\end{document}